\documentclass{elsart}
\usepackage{natbib,epsfig}
\begin{document}
\runauthor{Marco Battaglia}
\begin{frontmatter}
\title{A Study of Monolithic CMOS Pixel Sensors Back-thinning and their
Application for a Pixel Beam Telescope}
\author[UCB,LBNL]{Marco Battaglia}
\author[LBNL]{Devis Contarato}
\author[LBNL,INFN]{Piero Giubilato}
\author[LBNL]{Leo Greiner}
\author[UCB]{Lindsay Glesener}
\author[UCB]{Benjamin Hooberman}
\address[UCB]{Department of Physics, University of California at 
Berkeley, CA 94720, USA}
\address[LBNL]{Lawrence Berkeley National Laboratory, 
Berkeley, CA 94720, USA}
\address[INFN]{Istituto Nazionale di Fisica Nucleare, Sezione di Padova, Italy}

\begin{abstract}
This paper reports results on a detailed study of charge collection and 
signal-to-noise performance of CMOS monolithic pixel sensors before and 
after back-thinning and their application in a pixel beam telescope 
for the ALS 1.5~GeV $e^-$ beam test facility.

\end{abstract}
\end{frontmatter}

\typeout{SET RUN AUTHOR to \@runauthor}


\section{Introduction}

The Vertex Tracker for the International Linear Collider (ILC) has requirements
that largely surpass those of the detectors at LEP, SLC and LHC. Not only does 
the single point resolution need to be improved to just a few microns, but the 
material budget should not exceed $\sim$0.1\%~$X_0$ per layer. This is mainly
motivated by the need to efficiently identify all of the secondary particles 
in hadronic jets initiated by a heavy quark in order to distinguish $c$ from 
$b$ jets and also to determine the vertex charge. This translates to a 
requirement for Si pixel chips that do not exceed 50~$\mu$m in thickness. 
CMOS monolithic active pixel sensors are ionising radiation detectors featuring
a field-free, un-depleted sensitive volume. Because charge generation is 
confined primarily to a thin epitaxial layer of just 10-20~$\mu$m, it is 
possible to remove most of the bulk silicon using a back-thinning process 
without significantly affecting the signal charge collected. This makes CMOS 
detectors an appealing candidate for meeting the ILC physics requirements in 
terms of  material budget. There has been some successful experience in 
back-thinning full wafers of CMOS pixel structures, but questions on the effect
of back-thinning on charge collection arise~\cite{kleinfelder,dulinski}. 
In this study we address the issues of back-thinning individual diced chips, 
characterise the sensor performance before and after thinning to assess possible 
changes in signal charge collection and apply them in a beam telescope for use 
at a 1.5~GeV $e^-$ beam test facility.

\section{Back-thinning Studies}

The Mimosa~5 chip~\cite{mimosa, deptuch-ref}, developed at IPHC in 
Strasbourg, France, has been selected for this study. This chip,
fabricated using the 0.6~$\mu$m AMS process, features a large 
active area of 1.7$\times$1.7~cm$^2$ and more than 1~M pixels. 
The epitaxial layer is 14~$\mu$m thick and the pixel pitch 17~$\mu$m.
The wafer thickness is 450~$\mu$m. 
The detector is sub-divided into four independent sectors. For testing, 
the chip has been mounted on a mezzanine card which is installed on the 
readout board. 

Data acquisition and on-line analysis is performed by a LabView program, while 
the offline data analysis is performed by a dedicated C++ program. 
The chip characterisation procedure consists of three steps. The chip is 
mounted on the mezzanine card using a removable glue. First its charge-to-voltage 
conversion gain is measured using a $^{55}$Fe source, then charge generation at 
various depths is probed using collimated lasers of various wavelengths and 
finally the response to minimum ionising particles is studied on the 1.5~GeV 
$e^-$ beam of the LBNL Advanced Light Source (ALS). The chip is then removed 
by detaching the wire bonds and placing the chip and mezzanine card 
into a heated solvent bath and sent for back-thinning.
After back-thinning, the chip is visually inspected, permanently 
attached to the mezzanine card with film adhesive and re-bonded. 
The same characterisation procedure is repeated and the comparison of results 
allows us to determine the change in chip performance due to the back-thinning 
procedure.

The detector gain is determined using the 5.9~keV X-rays emitted by a 
$^{55}$Fe source. Charge generated by X-rays which convert in the shallow 
depletion region near the pixel diode is fully collected, resulting in a 
pulse height peak corresponding to the full X-ray energy, or 1640 electrons.
This provides the ADC count calibration in electrons. Typical values for the 
Mimosa~5 chip in our setup are in the range of 6 to 8 electrons per ADC count.
The collection of charge generated at different depths in the Si thickness
is studied using the fact that the penetration depth of IR photons in silicon 
is strongly dependent on their wavelength. We use an 850~nm laser to probe 
charge generation restricted primarily to the epitaxial layer and a 1060~nm 
which generates charge throughout 
the full wafer thickness. The setup consists of a laser diode pig-tailed to a 
6~$\mu$m-core optical fibre.  The fibre is terminated on an aspheric lens 
doublet which provides a collimated beam with a nearly Gaussian profile. 
Data is analysed by selecting a fixed 25$\times$25 pixel matrix centred 
around the signal maximum and integrating the charge collected in the matrix 
over several hundred events.  
The detector response to minimum ionising particles is determined using the 
1.5~GeV $e^-$ beam extracted from the booster ring of the ALS. The readout 
cycle consists of a reset followed by the readout of three subsequent frames 
of a 510$\times$512 pixel sector. The readout sequence is synchronised with the 
1~Hz booster extraction cycle so that the beam spill hits the detector just 
before the second frame. The empty frames are used to update pedestal and noise 
values. Correlated double sampling is performed on line.

Back-thinning is performed by Aptek Industries~\cite{aptek-ref} using a proprietary 
hot wax formula for mounting wafers and die to stainless steel grinding plates. 
The use of wax as an adhesive offers greater flexibility for handling thinner 
parts as well as eliminating the effects of ESD damage. The back-thinning is 
performed by a wet grind process with a rust inhibitor for cooling the chips 
and keeping the grind wheel free of debris which could cause damage when thinning 
below 100~$\mu$m. The process allows accurate thickness measurements in-situ. 
After the grinding a polish process is performed which minimises the stress from 
the backside of the device and allows to achieve thicknesses below 50~$\mu$m. 
Yields are dependent on various factors related to the quality of the silicon, 
including where in the ingot the wafers are taken from. Front-side processing 
factors such as oxides or polyamides as well as doping of a wafer can cause 
stress in the silicon lattice and may result in failure in the silicon at 
ultra thin specifications.
Four chips have been characterised and back-thinned as part of this study. 
Three have been thinned down to 50~$\mu$m while a fourth sensor has been 
thinned to 39~$\mu$m. At this thickness, chipping of the sensor edges was 
observed and the process was stopped. Inspection under a microscope revealed that 
the edges of the chip have been damaged, but only outside the guard ring area, 
thus not affecting the electrical functionalities of the device. Results 
reported here are based on the analysis of all four sectors of one chip 
thinned to 50~$\mu$m and of the chip thinned to 39~$\mu$m.

\begin{table*}
\caption{Summary of the results on back-thinned chip tests, given as average percentage 
changes of signal pulse heights after and before back-thinning, quote uncertainties are 
the rms values of the results for the different sectors.}
\label{tab:table1}
\begin{center}
\begin{tabular}{|c|c|c|c|c|c|}
\hline  \textbf{Thickness} & \textbf{Noise} & \textbf{$^{55}$Fe} & \textbf{850 nm} & \textbf{1060 nm} & \textbf{1.5~GeV $e^-$} \\ 
        \textbf{($\mu$m)}  &                &                    &                 &                  &          \\ \hline
50  & (+3$\pm$7)\%   &  (-7$\pm$8)\%  & (-16$\pm$6)\% & (-16$\pm$10)\%   & (-9$\pm$7)\%  \\ 
39  & (+8$\pm$13)\%  &  (+2$\pm$2)\%  & (-10$\pm$6)\% & (+130$\pm$42)\%  & (+2$\pm$4)\%  \\ \hline
\end{tabular}
\end{center}
\end{table*}

Results are summarised in Table~\ref{tab:table1} in terms of the relative change 
of the response before and after back-thinning. Values for the 
$^{55}$Fe data refer to the centroid of the 5.9 keV~peak, those for the laser 
measurements refer to the mean value of the pulse height distribution while 
those for the 1.5~GeV $e^-$ beam refer to the most probable value of a Landau 
function fitted to the reconstructed cluster pulse height distribution. 

The results of these tests show no significant degradation in the amount of 
collected charge after back-thinning to 39$~\mu$m and 50~$\mu$m. 
$^{55}$Fe data shows that the gain is not significantly affected. 
Instead laser data indicate that the collected charge varied, after processing, 
by an amount from -16\% to +130\% with the larger variations observed for longer 
wavelength. The laser output was monitored using a Si photo-detector and found 
to be stable to better than 10\%. The signal increase could then be explained as 
a change of the optical properties at the backplane after the thinning process. 
This effect will be further investigated, but it does not affect the sensor 
response to a minimum ionising particle.
The change observed for the lower wavelength, which is absorbed within the epitaxial 
layer, is consistent with the measurement uncertainty, due to the stability of the 
laser source. The ALS data confirm that neither the charge collection nor the 
signal-over-noise ratio of the detectors for energetic charged particles are 
affected by the back-thinning process down to thicknesses below 50~$\mu$m.

\begin{figure}
\begin{center}
\begin{tabular}{c c}
\epsfig{file=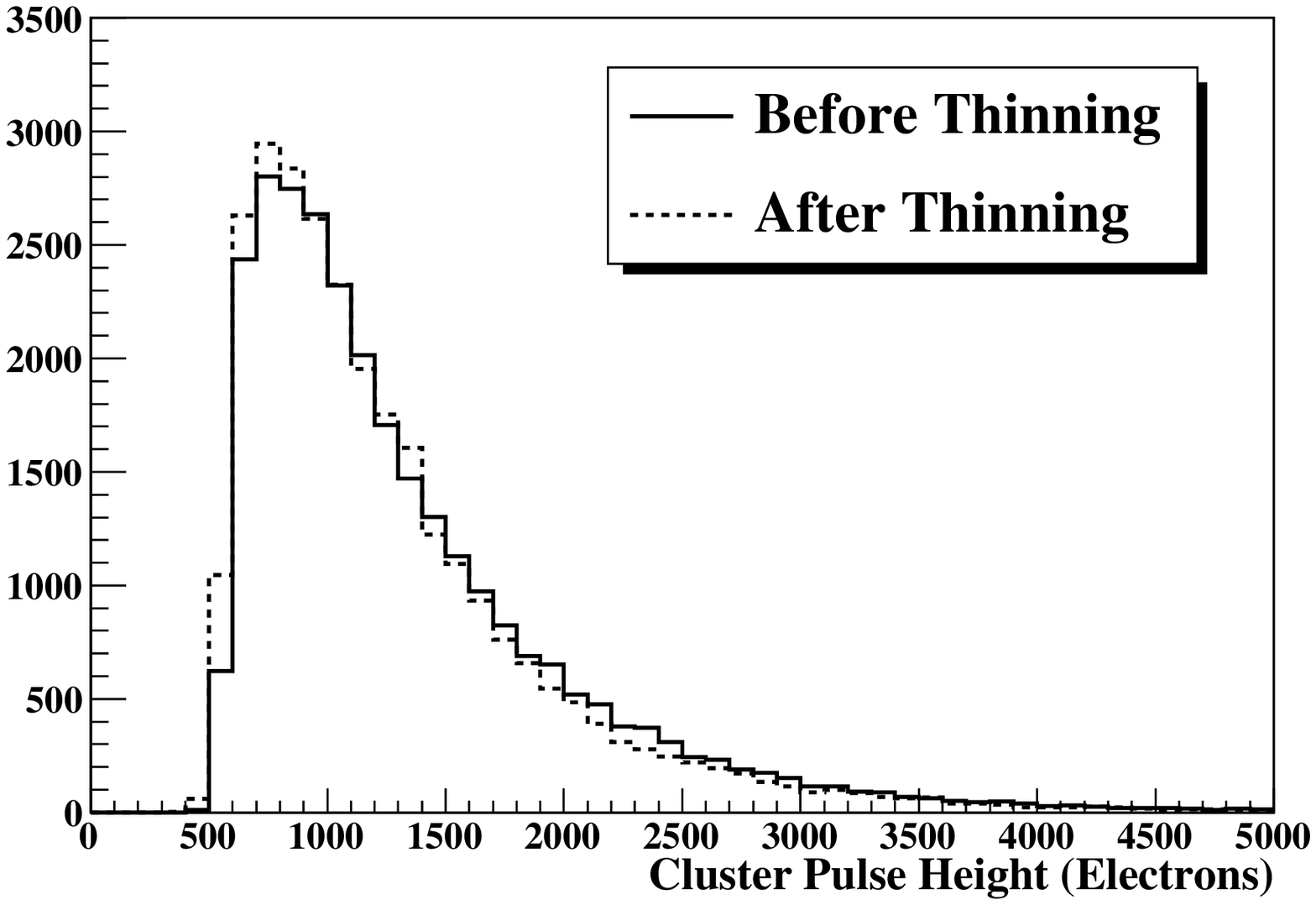,width=6.75cm}&
\epsfig{file=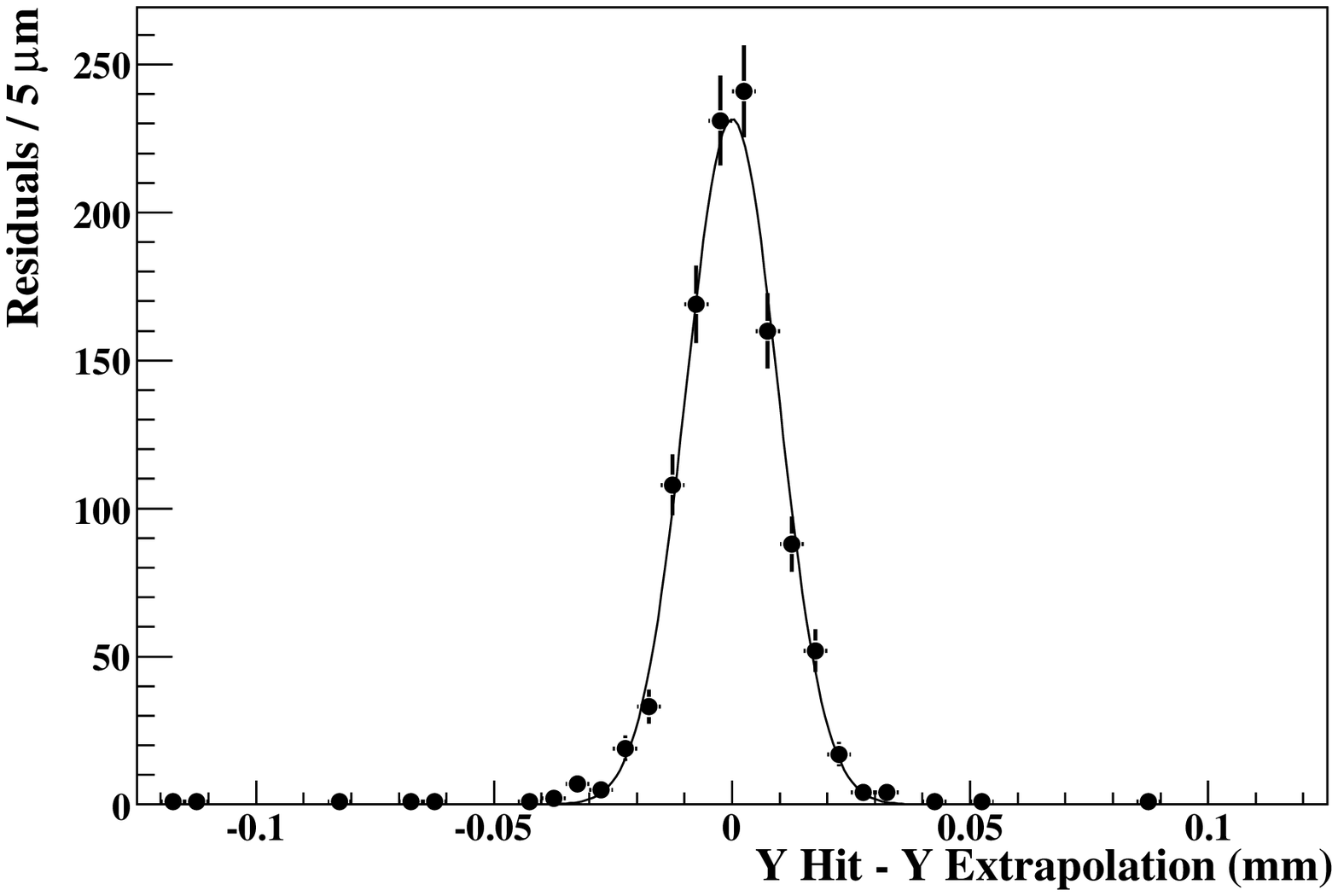,width=6.75cm}\\
\end{tabular}
\end{center}
\caption[]{Backthinning test (left): 1.5~GeV $e^-$ cluster pulse height, 
before (solid) and after (dotted line) backthinning. Beam telescope performance 
(right): residuals between pixel hits on the first layer and closest extrapolated 
track position for data collected in a single run on the ALS 1.5~GeV $e^-$ beam. 
The fitted gaussian curve has a width of 9.5~$\mu$m.}
\label{fig:fig1}
\end{figure}

\section{The CMOS Thin Pixel Pilot Telescope}

The batch of thinned Mimosa~5 chips has been used to build a CMOS Thinned 
Pixel Pilot Telescope (TPPT) installed on the 1.5~GeV $e^-$ ALS BTS 
facility at LBNL. The ongoing LBNL R\&D program on monolithic pixel sensors 
requires beam tests of sensor prototypes, which are routinely performed at 
the ALS. But the relatively low energy of the beam extracted from the ALS booster 
would make it difficult to track particles with enough accuracy to study single 
point resolution and efficiency of the detector under test, using conventional 
Si detectors. The TPPT provides a solution to this problem, offering at the 
same time an interesting prototype CMOS pixel tracker for detailed tracking 
performance studies towards the ILC Vertex Tracker. This is the first Si beam 
telescope made of thinned CMOS sensors and consists of three planes of thin pixel 
sensors (layers 1 to 3), each spaced by 17~mm. One additional Mimosa~5 chip (layer 4) 
is added 17~mm downstream of the last layer. This detector can be replaced by 
other detectors under test (DUT) with their independent readout electronics and 
placed as close as 5~mm from the latest telescope plane.
One sector of each chip, corresponding to a 510$\times$512 
pixel array, is readout through a custom FPGA-driven acquisition board. 
Four 14~bits, 40~MSample/s ADCs simultaneously read the four sensors, 
while an array of digital buffers drive all the
required clocks and synchronisation signals. The FPGA has been
programmed to generate the clocks pattern and collect the sampled data 
from the ADCs. A 32~bits wide bus connects the FPGA to a
digital acquisition board installed on a control PC. Data is processed on-line 
by a LabView-based program, which performs correlated double sampling and 
pedestal subtraction, noise computation and cluster identification.  To reduce 
the amount of data written to disk only the addresses and pulse heights of 
the pixels in a fixed matrix around the centre of a cluster candidate 
are recorded. The data is converted in the {\tt lcio} format for offline 
analysis using a set of dedicated processors developed for the {\tt Marlin} 
framework~\cite{Gaede:2006pj}. 
The telescope is operated in an optical enclosure mounted on 
an optical rail and aligned on the BTF beam line. The temperature is kept 
constant during operation at 27$^{\circ}$ by forced airflow. The measured 
average signal-over-noise for clusters associated to particle tracks is 14.4.

Detailed simulation of the TPPT has been carried out, using the
{\tt Geant-4} package~\cite{Agostinelli:2002hh} to generate the 
particle points of impact and energy deposits on the sensitive planes.
These have been stored in {\tt lcio} format and used as input to 
a CMOS pixel simulation program implemented in
{\tt Marlin}~\cite{elba06}. The simulation has been validated on 
beam test data collected with the Mimosa~5 chip on the 1.5~GeV ALS beam 
and on the 6~GeV $e^-$ DESY beam~\cite{devis}. 
Noise values have been matched to those measured for the detectors in 
the telescope. The simulated response has been passed to the same cluster 
reconstruction  program used for real data and reconstructed hits used to 
fit straight particle tracks, after pattern recognition. The telescope 
performance has been characterised by the residuals between the extrapolation 
of a track reconstructed on layers 2, 3 and 4 back to the first layer. 
Results from simulation have been used for optimising the TPPT geometry. 
By placing the DUT 5~mm away from the last telescope layer, an extrapolation 
resolution better than 4~$\mu$m can be achieved.

Data have been collected at the BTF with different beam intensities ranging 
from 0.5 particles~mm$^{-2}$ up to about 5 particles~mm$^{-2}$.
After data taking, the beam telescope geometry has been surveyed using an optical
metrology machine. The results of this survey have been used as starting point
of the alignment procedure, performed on a sample of approximately 20000 
well-isolated particle tracks with four correlated hits. After alignment, 
track candidates have been defined by matching hits in the second and fourth layer. 
Each track candidate has been extrapolated to the third, intermediate layer, where 
the closest hit has been added if its residual was less than 50~$\mu$m. 
The track candidate ambiguities have been solved based on the number of associated 
hits and on the difference between the track slope 
and the expected beam slope, determined from the settings of the beam-line final dipoles. 
Tracks have then been re-fitted using a modified least-square, to account for kinks due 
to multiple scattering on the measuring planes~\cite{Lutz:1987vs}, and extrapolated to 
the first layer where the residual to the closest reconstructed hit has been computed. 
Since the beam contains a fraction of lower momentum particles, which increases while 
moving away from the beam axis, only the central region of the beam was used and an 
additional cut was imposed on the residual on the second coordinate. A preliminary 
result for the residual gives 9.5~$\mu$m for all accepted tracks, which becomes 
9.1~$\mu$m when restricting to three-hit tracks. This should be compared to
6.8~$\mu$m obtained from the simulation, which assumes perfect geometry. 

\vspace*{-0.75cm}

\section*{Acknowledgements}

\vspace*{-0.75cm}

We thank Marc Winter and Frank Gaede for discussion and Robert Foglia, of Aptek 
Industries, for assistance with the back-thinning process. 
This work was supported by the Director, Office of Science, of the 
U.S. Department of Energy under Contract 
No.DE-AC02-05CH11231 and used resources of the National Energy Research Scientific 
Computing Center, supported under Contract No.DE-AC03-76SF00098.
We are indebted to the staff of the LBNL Advanced Light Source for their 
help and the excellent performance of the machine.

\end{document}